\documentstyle[12pt]{article}
\def\lsim{\:\raisebox{-0.5ex}{$\stackrel{\textstyle<}{\sim}$}\:}
\def\gsim{\:\raisebox{-0.5ex}{$\stackrel{\textstyle>}{\sim}$}\:}
\newcommand {\ignore}[1]{}

\newcommand{\noi}{\noindent}
\newcommand{\bc}{\begin{center}}
\newcommand{\ec}{\end{center}}
\newcommand{\epm}{e^+e^-}
\def\ifmath#1{\relax\ifmmode #1\else $#1$\fi}
\def\3quarter{{\textstyle{3 \over 4}}}

\def\vs{\vskip}

\def\ra{\rightarrow}

\def\lf{\leaders\hbox to 1em{\hss.\hss}\hfill}

\def\21{$SU(2) \ot U(1)$}
\def\321{$SU(3) \ot SU(2) \ot U(1)$}

\def\nt{\hbox{$\nu_\tau$ }}

\def\O{\hbox{$\cal O$ }}

\def\mnt{\hbox{$m_{\nu_\tau}$ }}

\def\neu{\hbox{neutrino }}

\def\eq#1{{eq. (\ref{#1})}}

\def\VEV#1{\left\langle #1\right\rangle}

\def\lsim{\raise0.3ex\hbox{$\;<$\kern-0.75em\raise-1.1ex\hbox{$\sim\;$}}}
\def\gsim{\raise0.3ex\hbox{$\;>$\kern-0.75em\raise-1.1ex\hbox{$\sim\;$}}}

\def\bel{\begin{letter}}
\def\eel{\end{letter}}
\def\beq{\begin{equation}}
\def\eeq{\end{equation}}
\def\bef{\begin{figure}}
\def\eef{\end{figure}}
\def\bet{\begin{table}}
\def\eet{\end{table}}
\def\bea{\begin{eqnarray}}
\def\ba{\begin{array}}
\def\ea{\end{array}}
\def\bi{\begin{itemize}}
\def\ei{\end{itemize}}
\def\ben{\begin{enumerate}}
\def\een{\end{enumerate}}
\def\ra{\rightarrow}
\def\ot{\otimes}

\def\eea{\end{eqnarray}}
\def\nps#1#2#3{          {\it Nucl. Phys. B (Proc. Suppl.) }
                         {\bf #1} (19#2) #3}
\def\np#1#2#3{           {\it Nucl. Phys. }{\bf #1} (19#2) #3}
\def\pl#1#2#3{           {\it Phys. Lett. }{\bf #1} (19#2) #3}
\def\pr#1#2#3{           {\it Phys. Rev. }{\bf #1} (19#2) #3}
\def\prep#1#2#3{         {\it Phys. Rep. }{\bf #1} (19#2) #3}
\def\prl#1#2#3{          {\it Phys. Rev. Lett. }{\bf #1} (19#2) #3}

\def\n.c.#1#2#3{         {\it Nuovo Cim. }{\bf #1} (19#2) #3}
\def\r.n.c.#1#2#3{       {\it Riv. del Nuovo Cim. }{\bf #1} (19#2) #3}

\def\ppnp#1#2#3{           {\it Prog. Part. Nucl. Phys. }{\bf #1} (19#2) #3}

\relax
\begin{document}
\begin{titlepage}
\rightline{FTUV/94-72}
\rightline{IFIC/94-69}
\rightline{hep-ph 9502237}
\noindent
\today
\begin{center}
{\Large \bf NOVEL SCALAR BOSON DECAYS IN SUSY WITH BROKEN R-PARITY }\\
\vskip .4cm
{\large \bf F. de Campos},
\footnote{E-mail CAMPOSC@evalvx.ific.uv.es}\\
{\large \bf M. A. Garc\'{\i}a-Jare\~no},
\footnote{E-mail GARCIAMA@evalvx.ific.uv.es}\\
{\large \bf Anjan S. Joshipura}
\footnote{E-mail anjan@prl.ernet.in}\\
{\large \bf J. Rosiek}
\footnote{E-mail ROSIEK@evalvx.ific.uv.es and ROSIEK@fuw.edu.pl}\\
and\\
\vs .1cm
{\large \bf Jos\'e W. F. Valle}
\footnote{E-mail VALLE@flamenco.ific.uv.es and VALLE@evalvx.ific.uv.es }\\
Instituto de F\'{\i}sica Corpuscular - IFIC/C.S.I.C.\\
Dept. de F\'{\i}sica Te\`orica, Universitat de Val\`encia\\
46100 Burjassot, Val\`encia, SPAIN\\
\vs .1cm
\end{center}
\vskip 1.0cm
\begin{abstract}
\baselineskip=14pt
{R parity violation can induce mixing of the supersymmetric
Higgs bosons with the sneutrinos at the tree level. We study
the effect of this mixing on the decays of Higgs scalars
as well as sneutrinos in an effective model where the violation
of $R$ parity is included in the minimal supersymmetric model
through bilinear lepton number violating superpotential terms.
We show that a small violation of $R$ parity can lead to a
sizeable branching ratio for the supersymmetric Higgs boson decay
mode $H \rightarrow \chi \ell$ (where $\chi$ denotes an
electroweak gaugino
and $\ell$ is either a tau neutrino or a tau lepton). Relevant
constraints on $R$ parity violation as well as those coming
from SUSY particle searches still allow the decay
$H \rightarrow \chi \ell$
to compete with the conventional decay $H\ra b \bar{b}$,
at least for some ranges of parameters of the model.
Moreover, the tau sneutrino will have dominant R parity
violating decays to standard model fermions
$b \bar{b}$, $\tau^+ \tau^-$ or to the invisible
mode $\nu \bar{\nu}$ whenever the phase space for
R parity conserving channels is closed.
}
\end{abstract}

\end{titlepage}

\section{Introduction}

The minimal supersymmetric standard model (MSSM) \cite{mssm}
provides an attractive generalization of the standard electroweak
theory and its phenomenology is quite well studied \cite{aleph}.
In analysing the  consequences of the MSSM
one usually assumes \cite{wein} the conservation of a discrete $Z_2$
symmetry, called R parity, distinguishing matter fields from their
superpartners. Given the importance of this assumption and
the fact that it does not follow on any general grounds,
it is important to analyse the possible implications of relaxing
it in phenomenological studies of supersymmetric models.
Considerable effort has gone into studying
consequences of R parity violating versions of the MSSM
\cite{hall}-\cite{asj2}, as well its extensions which realize
the spontaneous violation of R parity \cite{MASI_pot3,ROMA}.
The breakdown of R parity and that of the \21 symmetry can result in
the mixing of the ordinary particles with the superparticles
having the same electric charge and spin. Thus the R parity
violation could lead to mixing of ({\em i}) gauginos ($\lambda$)
with leptons ($\ell$) and ({\em ii}) Higgs bosons with sleptons
($\tilde{\ell}$).
The consequences of the gaugino-lepton mixings have been analysed
in many papers \cite{valleross,dp}. Such mixing leads to neutrino
masses \cite{hall,asj2,mas} and decays \cite{RPMSW}, as well as to
R parity violating Z decays such as $Z \rightarrow \lambda + \ell$
\cite{ROMA}.

In contrast to the $\lambda-\ell$ mixing, the consequences
of the mixing of Higgs bosons with sleptons are relatively less
explored \cite{mas1,asj1,asj2}. This mixing could have important
implications in the Higgs sector. For example, it was shown
recently in \cite{asj1}, that one could violate CP spontaneously
in the MSSM as a consequence of such mixing. Here we study an
important consequence of the $H-\tilde{\nu}$ mixing, namely,
the possibility of Higgs boson decays into gauginos and leptons.
These decays differ from the corresponding R parity violating
$Z$ decays in an important way.

Various experimental constraints on the magnitude
of R parity violation restrict the $\chi-\ell$ mixing and hence
the attainable values of the branching ratios for R parity
violating Z decays to be at the level of \O(10$^{-5}$) or so
\cite{ROMA} and therefore within the sensitivities of the LEP
experiments. Similar constraints also limit the amount
of mixing between the Higgs bosons and the sleptons. However,
in contrast to the Higgs boson, which mainly decays
through Yukawa couplings, the sleptons decay through
gauge interactions. This can compensate for the smallness
of the Higgs-slepton mixing, opening the possibility
for sizeable R parity violating Higgs boson decays.

In fact, as we will show explicitly, if allowed kinematically,
the SUSY decay e.g. $H \rightarrow \chi \ell$ ($\chi$ being here
lightest neutralino, often assumed to be also the lightest
supersymmetric particle - or LSP, and $\ell$ being in this
case the \nt) can become comparable or even dominate over the
conventional decay to $b \bar{b}$ pair for some ranges of
SUSY parameters.

The $H-\tilde{\nu}$ mixing can also have sizeable
effects on the sneutrino decay pattern as well.
If, as usually assumed in the MSSM, one of the neutralinos
is the lightest supersymmetric particle, the sneutrinos
will decay to the LSP and a neutrino. If not, then the
sneutrinos will only have three body decay modes.
In contrast, in the present case, if all neutralinos
are heavier than the sneutrino, the latter would dominantly
decay to non supersymmetric channels, for example $b \bar{b}$
pairs (through the $H-\tilde{\nu}$ mixing) or to $\nu\bar{\nu}$ pairs
(through $\chi^0-\nu$ mixing). Similar decays to tau pairs would
also be induced by $\chi-\tau$ mixing.
The existence of such new sneutrino decay modes can significantly
influence the conventional strategy for sneutrino search.

In this note, we explore the interesting range of the allowed
parameter space of the MSSM for which the decay
$H \rightarrow \chi \ell$ is comparable to the
standard decay mode $H\ra b \bar{b}$ and discuss
the possible signatures associated with this SUSY
decay mode. We also show that, quite generally,
the R parity violating sneutrino decay channels
to standard model fermions are dominant below
the threshold for R parity conserving supersymmetric
decay channels.

\section{Basic Framework}

The MSSM is characterised by the following superpotential,
written in standard notation:
\beq \label{w0}
W_0 = \varepsilon_{ab}\left [h_{ij}\hat{L}_i^a \hat{H}_1^b
\hat{E}_j^C + h'_{ij}\hat{Q}_i^a \hat{H}_1^b \hat{D}_j^C + h''_{ij}
\hat{Q}_i^a \hat{H}_2^b \hat{U}_j^C + \mu \hat{H}_1^a \hat{H}_2^b \right]
\eeq
One could also add to this the following R violating terms
\beq \label{wr}
W_R=\varepsilon_{ab}\left[\lambda_{ijk}\hat{L}_i^a \hat{L}_j^b
\hat{E}_k^C + \lambda'_{ijk}\hat{L}_i^a \hat{Q}_j^b \hat{D}_k^C
+ \epsilon_i \hat{L}_i^a \hat{H}_2^b \right]
\end{equation}
where we have omitted the baryon number violating terms \cite{wein}
whose presence along with terms in \eq{wr} would lead to fast proton
decay.

In what follows we will focus only on the effect of the
bilinear R parity violating term in \eq{wr}, which is the only one
to have a direct effect on the physics of the neutral Higgs boson
sector. This term, characterised by strength parameters $\epsilon_i$,
plays a crucial role in generating Higgs-slepton and gaugino-lepton
mixing at the tree level.

The scalar potential contains the standard soft supersymmetry breaking
as well as supersymmetric  terms  following from the superpotential
in \eq{w0} and \eq{wr}. The neutral part of the potential can be written as
\bea \label{vhiggs}
V_{Higgs} & = &m_1^2 \vert \phi_1 \vert^2 + m_2^2 \vert \phi_2 \vert^2
+ m_{L_i}^2 (\tilde{\nu}_i^{\dagger}\tilde{\nu}_i)
+ \lambda \left( \vert \phi_1 \vert^2 - \vert \phi_2 \vert^2
+ \tilde{\nu}_i^{\dagger}\tilde{\nu}_i \right)^2 \nonumber \\
&+& \left[ \mu \epsilon_i  \phi_1 \tilde{\nu}_i^*+
B_1 \mu m_{3/2}\phi_1 \phi_2+
B_2 \epsilon_i m_{3/2} \phi_2\tilde{\nu}_i + h.c.\right]
\eea
where $\phi_{1,2} \equiv H_{1,2}^0$
and $\lambda=\frac{1}{8}(g^2+g'^2)$, with $g$ and $g'$ denoting
the $SU(2)$ and $U(1)$ couplings.
 The terms involving  parameters $\epsilon_i$ are generated by the
R violating terms in eq.(2). As a result of these terms, sneutrino
fields invariably acquire a nonzero VEV \cite{hall,mas1,arca}. For
definiteness we assume that only $\epsilon_3$ is non-zero
\footnote{By neglecting $\epsilon_1$  and $\epsilon_2$ we are
safe from the point of view of possible baryogenesis constraints
\cite{dreineross}}.

Above the electroweak (and supersymmetry) breaking scale
it is always possible to redefine the $\hat{L}_3$ and $\hat{H}_1$
superfields so as to eliminate the bilinear term
$\hat{L}_3 \hat{H}_2$ from the superpotential. However this does not
mean that an analogous term is not present in the
scalar potential. Indeed, the redefinition of $\hat{L}_3$
and $\hat{H}_1$ through the orthogonal transformation
\begin{eqnarray}
\hat{H}_1' = {\mu \hat{H}_1 + \varepsilon_3 \hat{L}_3\over\sqrt{\mu^2
+\varepsilon_3^2}}
\nonumber\\
\hat{L}_3' = {-\varepsilon_3 \hat{H}_1 + \mu \hat{L}_3\over\sqrt{\mu^2
+\varepsilon_3^2 }}
\nonumber\\
\end{eqnarray}
does not leave the scalar supersymmetry breaking mass
terms invariant, since they are not expected to be
universal below the unification scale. This implies
that, after the substitution made in order to
remove the $\hat{L}_3\hat{H}_2$ term from superpotential,
one generates a linear term in the scalar potential
for the slepton field, as a result of which a non-zero
VEV for the scalar neutrino is induced.

The potential in \eq{vhiggs} is minimized for
\bea
v_1[m_1^2+\lambda c]+B_1 \mu m_{3/2} v_2+\mu \epsilon_3 v_3&=&0
\nonumber\\
v_2[m_2^2-\lambda c]+B_1 \mu m_{3/2} v_1+B_2 \epsilon_3 m_{3/2} v_3&=&0
\nonumber\\
v_3[m_{L_3}^2+\lambda c]+\epsilon_3 [B_2  m_{3/2} v_2+\mu  v_1]&=&0
\eea
where $\VEV{\phi_1} \equiv \frac{v_1}{\sqrt 2}$,
$\VEV{\phi_2} \equiv \frac{v_2}{\sqrt 2}$,
$\VEV{\tilde{\nu_3}} \equiv \frac{v_3}{\sqrt 2}$ and
$c \equiv (v_1^2 - v_2^2 + v_3^2)$.
The presence of a non-zero $v_3$ leads to mixing of the scalar
(pseudoscalar) Higgs with the scalar (pseudoscalar) component
of the sneutrino. The relevant mass matrices are now given by
\beq
M_R^2=\left(
\ba{ccc}
m_1^2 + \lambda c+ 2 \lambda v_1^2& -\lambda v_1 v_2+B_1 \mu m_{3/2}
& \lambda v_1 v_3+\mu \epsilon_3\\
- \lambda v_1 v_2+B_1 \mu m_{3/2} & m_2^2-\lambda c+2 \lambda v_2^2 &
- \lambda v_3 v_2+B_2 \epsilon_3 m_{3/2}\\
 \lambda v_1 v_3+ \mu \epsilon_3 & - \lambda v_3 v_2+B_2 \epsilon_3 m_{3/2} &
m_{L_3}^2+\lambda c+2 \lambda v_3^2\\
\ea \right)
\eeq

\beq
M_I^2=\left(
\ba{ccc}
m_1^2+\lambda c&-B_1 \mu m_{3/2}&
\mu \epsilon_3\\
-B_1 \mu m_{3/2}&m_2^2-\lambda c&
-B_2 \epsilon_3 m_{3/2}\\
 \mu \epsilon_3&-B_2 \epsilon_3 m_{3/2}&
m_{L_3}^2+\lambda c\\
\ea \right)
\eeq
The $\epsilon_3$-dependent terms in the above equation lead to
Higgs-sneutrino mixing and causes the R parity violating Higgs bosons decay
modes. In order to see the effects of these decay  modes note that
the decay of Higgs to $\chi$ and $\nu$ occurs either through the
sneutrino component of the lightest Higgs boson $\equiv h$,
\beq
h=a_{31} (\tilde{\nu})_R + a_{21} (\phi_2)_R+a_{11} (\phi_1)_R
\eeq
(where the subscript $R$ denotes the real part), or through the
\nt admixture in the lightest neutralino in the $h \chi\chi$ vertex.

The relative importance of the SUSY decay mode
$h \ra \chi \nu$ follows from the ratio
\begin{eqnarray}
\label{R}
R_0& \approx &\frac{\Gamma(h\ra \chi \nu)}{\Gamma(h\ra b \bar{b})}=
\frac{\tan^2 \theta_W}{2}\left(\frac{M_W}{m_b}\right)^2
\left(\frac{ (1- m_{\chi}^2/M_H^2)^2 }{ (1-4 m_b^2/M_H^2)^{3/2} }\right)
\frac{a_{31}^2}{a_{11}^2} \cos^2\beta \vert \xi \vert^2
\end{eqnarray}
where $\xi$ denotes the appropriate gaugino mixing factor
specifying the direction of the LSP. Although less likely
kinematically, one may also have decays of supersymmetric
Higgs bosons into charginos and taus. The relative
importance of this decay mode follows from the ratio
\begin{eqnarray}
\label{R+}
R_+& \approx
&\frac{\Gamma(h\ra \chi^+ \tau^- + \chi^- \tau^+)}{\Gamma(h\ra b \bar{b})}=
\frac{2}{3}\left(\frac{M_W}{m_b}\right)^2
\left(\frac{ (1- m_{\chi}^2/M_H^2)^2 }{ (1-4 m_b^2/M_H^2)^{3/2} }\right)
\frac{ a_{31}^2 }{ a_{11}^2 } \cos^2\beta \vert \xi^{\prime } \vert^2
\end{eqnarray}
where $ \xi^{\prime}$ is the wino-lightest chargino mixing element.

Similarly one can induce tau sneutrino decays to normal
standard model fermions due to R parity breaking
either in the scalar sector or in the chargino or
neutralino sectors. As we will discuss later,
the tau sneutrino will have dominant R parity
violating decays to standard model fermions
whenever the phase space for R parity conserving
channels is closed.

In determining these branching ratios one has also to
identify mass eigenstate charged and neutral fermions,
which follow from the corresponding neutralino and
chargino mass matrices. First we give the neutralino
matrix, expressed in the basis $-i \tilde{B}$,
$-i \tilde{W_3}$, $\tilde{h_1^0}$, $\tilde{h_2^0}$, $\nu_\tau$:
\beq
\label{M0}
M_{\chi^0}=\left(
\ba{ccccc}
xM&0&-\frac{g' v_1}{2}&\frac{g' v_2}{2}&-\frac{g' v_3}{2}\\
0&M&\frac{g v_1}{2}&-\frac{g v_2}{2}&\frac{g v_3}{2}\\
-\frac{g' v_1}{2} & \frac{g v_1}{2} & 0 & -\mu & 0\\
\frac{g' v_2}{2}&-\frac{g v_2}{2}&-\mu&0&\epsilon_3\\
-\frac{g' v_3}{2}&\frac{g v_3}{2}&0&\epsilon_3&0\\
\ea \right)
\eeq
where $M$ denotes the gaugino mass and we assumed the GUT
unification hypothesis $x= \frac{5 g'^2}{3 g^2}
= \frac{5}{3} \tan^2\theta_W \sim 0.5$

For charginos one has the following mass matrix $M_{\chi^\pm}$,
written in the basis where the rows correspond to
($-i \tilde{W^-}, \tilde{h_1^-}, \tau_L^-$)
while the columns stand for
($-i \tilde{W^+}, \tilde{h_2^+}, \tau_R^+$):
\beq
\label{M+}
M_{\chi^\pm}=\left(
\ba{ccc}
M & \frac{g v_1}{\sqrt{2}} & \frac{g v_3}{\sqrt{2}} \\
\frac{g v_2}{\sqrt{2}} & \mu & -\epsilon_3 \\
0 & \frac{h_{33} v_3}{\sqrt{2}}& \frac{-h_{33} v_1}{\sqrt{2}}\\
\ea \right)
\eeq
{}From the above formulas one expects that a modest mixing
($\sim$20\%) between the Higgs boson and the sneutrino can
give rise to a sizeable branching ratio for R parity violating
Higgs bosons decay modes. The same phenomena may also occur in
the decays of the pseudoscalar Higgs bosons. This will happen for
moderate $\epsilon_3$ values (which correspond to phenomenologically
acceptable \nt masses), if the {\sl sneutrino} mass is close
to the mass of the {\sl Higgs} boson, where by {\sl sneutrino}
we mean a state that is more than 50\% along the original weak
basis sneutrino, with the corresponding labeling for the Higgs case.
As we will demonstrate, it is indeed possible to obtain the required
amount of mixing satisfying all the relevant constraints on the
magnitude of the $R$ parity breaking that follow from experiment.

\section{Constraints}

We now turn to a more detailed analysis which includes all
relevant constraints on the parameters. The major constraint on
$\epsilon_3$ and hence on R parity violating mixings comes
from the neutrino mass. Both the non-zero  $\epsilon_3$ and
$v_3$ induce the mass for the $\nu_{\tau}$. This follows from
the structure of the neutralino mass matrix $M_{\chi}$ in \eq{M0}.
This matrix is of the seesaw form and results in
the following mass for the neutrinos \cite{asj2}
\beq
\label{mnt}
\mnt \approx \frac{xg^2+g'^2}{2\mu}\;\;\frac{(\mu v_3 + v_1 \epsilon_3)^2}
      {[-2x \mu M - v_1 v_2(xg^2+g'^2)]}
\eeq
Since \mnt is constrained to be $\leq 31\:$ MeV \cite{PDG94},
$\epsilon_3$ is correspondingly restricted.
The constraints on $R$ parity conserving parameters
come from the non-observation of the SUSY particles at LEP1.

Using the minimization equation, one could express all the elements
of the various mixing matrices in terms of six independent parameters
which we choose as
$\tan \beta = \frac{v_2}{\sqrt{v_1^2 + v_3^2}}$,
the $\mu$ parameter, the pseudoscalar mass parameter
$m_A^2 \equiv m_1^2+m_2^2$, the gaugino mass parameter $M$,
the soft sneutrino mass parameter $m_{L_3}$
\footnote{For simplicity we choose the parameter
$\gamma\equiv \frac{B_2}{B_1}$ to be 1},
and, finally, the parameter $\epsilon_3$. This last
parameter $\epsilon_3$ is absent in the standard
$R$ parity conserving MSSM case. As we will see,
this parameter is directly constrained by the tau
neutrino mass.

In our present analysis we have taken into account
the following constraints on the model parameters:
\begin{itemize}
\item
The tau neutrino mass bound following from the direct
searches in the laboratory \cite{PDG94}. Requiring
\mnt to be $\leq 31 MeV$ restricts the parameters in
\eq{mnt}.
On the other hand, the cosmological arguments based on the
closure density impose much stronger constraints \cite{KT}.
Indeed, a stable tau neutrino heavier than about 50 eV is
cosmologically forbidden. However, this problem
is easily solved in models where the violation of R parity
is spontaneous, since in this case the theory contains a
Goldstone boson, called majoron \cite{CMP,fae}. The \nt is then
unstable, decaying to a lighter neutrino plus majoron \cite{774,V}
and thereby avoiding an excessive relic \neu abundance \cite{RPMSW}.
Another cosmological constraint follows from
big bang nucleosynthesis \cite{BBNUTAU}.
However it seems reasonable to consider the
less conservative limit of $31$ MeV on \mnt,
since, even in the absence of a majoron, the off-diagonal
coupling of the $Z$ to \nt and a lighter neutrino can lead to
a cosmologically safe decay for the \nt into three neutrinos
\cite{774,2227}. Note that such couplings would arise in the
present case once the parameters $\epsilon_{1,2}$ (assumed
here to be zero for simplicity) are turned on. Although much
less efficient than the majoron decay channel, this decay is
enough for the case of large \nt masses, where the R parity
violating effects discussed here can be sizeable.
 \item
The lightest of the charginos should be heavier than $\sim 45$ GeV
as required by present negative result of supersymmetric searches at LEP.
 \item
Z decays to chargino and neutralinos may affect the partial Z decay
widths, which should obey the present restrictions imposed by LEP.
For the case of charginos this constraint follows automatically
from to the previous one, while for the neutralino case it depends
on SUSY parameter values
\end{itemize}
We have scanned the region of parameter space which is consistent with the
above constraints and for which the decay of supersymmetric Higgs bosons
to LSP and neutrino ($H \ra \chi^0 + \nu_\tau$) is kinematically allowed.
For definiteness, we have restricted our considerations to the
case of CP even Higgs bosons. The R parity violating decays of
pseudoscalar Higgs bosons $A \ra \chi^0 + \nu_\tau$ and
$A \ra \chi^{\pm} + \tau^{\mp}$ can be investigated in a
similar way.

The mixing $a_{31}$ appearing in \eq{R} and the analogous
one for the pseudoscalar case will become significant if
the {\sl sneutrino} mass is close to the mass of the
relevant {\sl Higgs} boson.

\section{Results }

Our results for R parity violating supersymmetric Higgs
boson decays are summarized explicitly in figures
1 to 6.

In fig. 1 we display the branching ratios
for CP even Higgs decays to LSP plus neutrino as
a function of the relevant Higgs boson-sneutrino
mass difference, for a suitable choice of SUSY
parameters, specified in the figure. In fig. 2
we display the corresponding "standard" $b \bar{b}$ decay
branching ratio for the same illustrative
choice of parameters. Clearly, for relatively
small Higgs boson-sneutrino mass differences
of a few GeV or so, the supersymmetric channel
can dominate over the standard one.

Conversely, it is possible in our model to substantially
affect sneutrino decay patterns, as the tau sneutrino
may decay into R parity violating standard model channels
such as $b \bar{b}$, $\tau^+ \tau^-$ or the invisible mode
$\nu \bar{\nu}$.
Indeed, these decays are the dominant ones whenever
the phase space for the R parity conserving channels
such as $\chi \nu$ is closed.
This is seen from figures 3, 4 and 5. For example,
for the choice of parameters given in fig. 3 the mass
of the lightest chargino is around 120 GeV. This explains
the drop in the branching ratio for sneutrino to
$\nu \bar{\nu}$ at this value of the sneutrino mass as,
above this threshold, the supersymmetric decay channel
would be open. This illustrates how the sneutrino decay
branching ratios into non-supersymmetric channels can be
sizeable when the supersymmetric channels are kinematically
forbidden.
On the other hand, fig. 4 illustrates that there
is a resonant enhancement of the $b \bar{b}$ and
$\tau^+ \tau^-$ decay branching ratios for
sneutrino masses close to the lightest CP even Higgs
boson mass.

The same effects can also be seen from fig. 5,
which corresponds to the $\tan \beta = 10$,
$M_2 = 70$ GeV, $\mu = 200$ GeV, $\epsilon_3 = 1$
and $m_A = 250$ GeV. Clearly, for sneutrino masses
below the threshold for LSP production (about 60 GeV
in this case)
the R parity violating standard model channels such
as $b \bar{b}$ are dominant. Moreover, they may be
nonnegligible even above the supersymmetric
threshold, provided the sneutrino masses lies
close to the lightest CP even Higgs boson mass,
leading to a resonant enhancement of the
$b \bar{b}$ or $\tau^+ \tau^-$ decay branching
ratios as discussed above. For the parameter set
used in fig. 5 this corresponds to $m_h= 155$ GeV
and $m_{\tilde{\nu}_{\tau}} \approx 150$ GeV and
leads to the small rise which can be seen from
this figure.
Note also that, for tau sneutrino masses below the
supersymmetric threshold, although the sneutrino will
be the LSP, it clearly will be unstable, since
R parity is broken, leading to visible signals
(such as a $b \bar{b}$ pair), not necessarily
missing momentum. These novel features may be
quite important in designing strategies to
search for the sneutrinos in our model.

As a final remark to close this section
we note that there is a wide class of
processes which exhibit R parity violation outside
Higgs boson sector \cite{granada}. For definiteness,
we calculate in this model, as an example, the
corresponding R parity violating Z decay
branching ratios \cite{ROMA}. We
focus on the most characteristic one, namely
single chargino production in Z decays. In fig. 6
we display the corresponding branching ratios
for Z decays to the lightest chargino plus
a tau of either charge, as a function of the
relevant \nt mass, after varying the SUSY
parameters over a reasonable range, specified as
$-250\:GeV<\mu<250\:GeV$ and $30\:GeV<M_2<200\:GeV$
and fixing the remaining relevant parameters
as follows:
$\tan \beta\;=5$, $m_{A}\;=50\; $ GeV,
$m_{\tilde{\nu}}\; =200\; GeV$, $\gamma\;=1$.
Clearly, in this model these branching ratios are
within the reach of the LEP experiments, for the
same range of $\epsilon_3$ or \mnt masses for which
the novel scalar boson decays modes suggested here
are also sizeable.

\section{Discussion }

We have demonstrated in this paper that for some ranges of
parameters of the MSSM model, the supersymmetric decay modes
of the Higgs bosons can be comparable to their conventional decays.
This feature can influence the supersymmetric Higgs boson
search strategies in a substantial way. Note that the R parity
breaking terms responsible for Higgs-sneutrino mixing also lead
to the decay of the LSP $\chi$. Therefore the neutralino emitted in
Higgs decays would be unstable and decay inside the detector if
$\epsilon_3 \sim $ few GeV. The LSP decay modes in this case will
consist of lepton + fermion-anti-fermion pairs \cite{ROMA}.
For example, the R parity violating Higgs boson decays
to the LSP, followed by the charged or neutral current
mediated LSP decays would give rise to
(a) two jets + missing momentum,
(b) di-lepton + missing momentum, or
(c) $\tau-e$ or $\tau-\mu$ pairs accompanied by missing
momentum signatures.
Folding now with the standard decays of the virtual $Z^0$
one gets the following signatures in $\epm$ collisions:
(i) two jets + missing momentum,
(ii) di-lepton + missing momentum,
(iii) di-lepton plus di-jets + missing momentum,
(iv) four jets + missing momentum,
(v) four leptons + missing momentum,
(vi) $\tau^+ \tau^-$ pairs plus two jets, and
(vii) $\tau-e$ or $\tau-\mu$ pairs accompanied by missing momentum.
Some of these signals, for example (vii), do not occur in the
standard model and, therefore, may be regarded as
characteristic of our supersymmetric decay modes.
It also does not appear in the MSSM, thus it appears
to be a characteristic feature of the broken R parity model.
For a sizeable branching ratio of $H\ra \chi \nu$,
these signals may be sizeable. Moreover, there may be,
for some choice of parameters, the possibility of Higgs
boson decays to charginos, which may also lead to novel
signatures.

We have performed our study in the framework of
the simplest model with R parity violation introduced in
an explicit way, aware of the fact that this may face
some stringent restrictions from nucleosynthesis, which
would be interesting to evaluate in detail, taking into
account the neutral current mediated decay of the \nt
into three neutrinos. We feel justified in using this
as a reasonable working model, since the novel decays
discussed here are also expected in a very general
class of models with spontaneous violation of R parity,
in which the cosmological restrictions are easily
evaded due to the majoron decays or annihilations
of the tau neutrino \cite{RPMSW}.

Our study opens the issues of how the existence of these novel
scalar decay channels affect the bounds on the Higgs masses
from the LEP experiments and of how they can influence the
search strategies for  Higgs bosons or sneutrinos at the
higher energies which will be available at LEP200
or LHC. We plan to return to these issues elsewhere.

Finally, we re-stress that striking R parity violating
effects are also expected to occur in the electroweak gauge
currents, leading to sizeable R parity violating Z decays,
as illustrated in fig. 6. Similary, R parity violating processes
could be visible at hadron colliders such as the upcoming
generation of Fermilab experiments and the LHC.

\section*{Acknowledgements}
\hspace{0.3cm} This work has been supported by DGICYT under
Grant numbers PB92-0084 and SAB94-0014 (A.S.J.) as well as a
postdoctoral fellowship (J.R.) and an FPI fellowship (M.A.G.J.).
The work of F. de Campos was supported by CNPq (Brazil).

\newpage

{\large \bf Figure Captions}

{\bf Figure 1}\\
\noi
Branching ratio of the lightest Higgs boson into
$\chi\;\nu_{\tau}$ as a function of the difference of
the lightest Higgs boson and the sneutrino mass.
We have fixed the supersymmetric parameters
$\mu$, $\tan\beta$, $M_2$ and the pseudoscalar mass $m_A$
as shown in the figure. The \nt mass was varied
as indicated by the different curves in the figure.

{\bf Figure 2}\\
\noi
Branching ratio of the lightest Higgs boson into
$b\bar{b}$ as as a function of the difference of
the lightest Higgs boson and the sneutrino mass.
The values of the parameters as well as the \nt
mass are exactly as in the previous figure.

{\bf Figure 3}\\
\noi
Branching ratio of the sneutrino into $\nu\bar{\nu}$
as a function of the sneutrino mass for five
different values for the tau neutrino mass
with the supersymmetric parameters fixed
as indicated in the figure.

{\bf Figure 4}\\
\noi
Branching ratio of the sneutrino into $b\bar{b}$
as a function of the difference of the lightest Higgs
boson and the sneutrino mass. Parameters are chosen as
in fig. 3.

{\bf Figure 5}\\
\noi
Branching ratio for sneutrino decays to
R parity breaking channels and supersymmetric channels,
as a function of the sneutrino mass. Parameters are
chosen as indicated in the text.

{\bf Figure 6}\\
\noi
$Z\rightarrow\; \chi\;\tau$ branching ratio
for different values of the lightest chargino
and \nt masses, with the supersymmetric parameters
chosen as described in the text. The area under
each line is the allowed region giving a branching
ratio in excess of the indicated value. All points
satisfy the observational constraints that follow
from SUSY particle searches as well as \neu physics
(see text).

\end{document}